\documentclass[12pt]{article}

\usepackage{graphicx}
\usepackage{graphics}
\usepackage{epsfig}
\usepackage{amssymb}
\usepackage{amsmath}
\usepackage{color}
\usepackage{textcomp}
\usepackage{latexsym}

\usepackage[margin=2.5cm,footskip=0.6cm]{geometry}
\begin{document}

\begin{center}
{\Large{\bf Single Particle Brownian Motion with Solid Friction}} \\
\ \\
\ \\
by \\
Prasenjit Das$^1$, Moshe Schwartz$^{2,3}$ and Sanjay Puri$^1$\\

$^1$School of Physical Sciences, Jawaharlal Nehru University, New Delhi 110067, India. \\
$^2$Beverly and Raymond Sackler School of Physics and Astronomy, Tel Aviv University, Ramat Aviv 69934, Israel.\\
$^3$Faculty of Engineering, Holon Institute of Technology, Golomb 52 Holon 5810201, Holon, Israel.
\end{center}

\begin{abstract}
\noindent We study the Brownian dynamics of a solid particle on a vibrating solid surface. Phenomenologically, the interaction between the two solid surfaces is modeled by solid friction, and the Gaussian white noise models the vibration of the solid surface. The solid friction is proportional to the sign of relative velocity. We derive the Fokker-Planck (FP) equation for the time-dependent probability distribution to find the particle at a given location. We calculate analytically the steady state velocity distribution function, mean-squared velocity and diffusion coefficient in $d$ dimensions. We present a generic method of calculating the autocorrelations in $d$ dimensions. This results in one dimension in an exact evaluation of the steady state velocity autocorrelation. In higher dimensions our exact general expression enables the analytic evaluation of the autocorrelation to any required approximation. We present approximate analytic expressions in two and three dimensions. Next, we numerically calculate the mean-square velocity and steady state velocity autocorrelation function up to $d=3$. Our numerical results are in good agreement with the analytically obtained results.
\end{abstract}
\newpage
\section{Introduction}\label{10sec1}
One of the cornerstones on which nonequilibrium statistical physics is based is the Langevin equation, phenomenologically describing the motion of a large particle interacting with a system of small interacting particles~\cite{balki}. While the noise affecting it is somewhat difficult to measure directly, the drag force, which is proportional to the velocity of the particle is easily accessible to macroscopic measurements. Another frictional force, perhaps even more important in most human activities is solid friction between touching macroscopic solid objects~\cite{LDV,GAM,CADC}. The main difference between the two is the way in which they depend on the velocity. While the drag force is proportional and opposite in direction to the velocity, the solid friction force is just a constant and points in the opposite direction of the velocity. The problem of solid friction is a very old problem, arises in many contexts, ranging from geology and engineering to physics and biology and to study the dynamics of a mesoscopic system~\cite{hokacc98,bnjp1998,jcjcrh00,ejb2002}. Solid friction is necessary in the study of stick-slip motion ~\cite{fje1997,rdal1998,feng2003,grp2008}, dynamics of dense granular fluids~\cite{rbms10,msrb11,pdspms}, vibrated granular media~\cite{ymms1998,akhh2004}, pouring of granular particles on an inclined plane and through a chute~\cite{gpeng9495,omnkmmhh,eafcpm99} as well as ratchet motion of a solid object on a vibrated surface~\cite{abfbpg06,dfyamu07}. 

There exist experimental studies where solid frictional interaction plays the crucial role. Goohpattader {\it et al.}~\cite{psg2009} studied the dynamics of a small object under solid friction in the presence of bias and white noise. They have calculated the displacement distribution experimentally and numerically. They found that the distribution function is non-Gaussian and that the variance of the distribution increases with time. Later, Gnoli {\it et al.}~\cite{agepl13,agprl13} studied the dynamical properties of a granular rotor subjected to dry friction and random excitation in a fluidized stationary granular gas. In their study, the experimental results for velocity distribution of the probe, autocorrelation function, and power spectra are in good agreement with the analytical predictions.

The problem of Brownian motion with solid or dry friction has also been studied extensively by using analytical techniques~\cite{amauger06,ycabhtwj,tskkhh16}. Dubkov {\it et al.}~\cite{dubkov09} studied the phenomenological Langevin equation for the particle motion in the presence of solid friction and interacting with a non-Gaussian thermal bath. Talbot {\it et al.}~\cite{talbot11} studied the motion of a granular rotor immersed in a bath of thermalized particles. A frictional torque acts on the axis of the rotor. Numerically, they found two scaling regimes. In the large friction limit, the asymptotic behavior of the Boltzmann-Lorentz equation describes the dynamics while in the limit of small friction and large rotor mass, the dynamics are described by the Fokker-Planck (FP) equation. Baule and Sollich~\cite{abps2012} studied an exactly solvable nonlinear model to describe the directed motion of an object in the presence of dry friction and shot noise with zero average. In their study, the object shows a unidirectional motion above a critical dry friction strength, resulting in a singular stationary velocity distribution. Sano and Hayakawa~\cite{tshh2014} studied the motion of an adiabatic piston under solid friction in a model system. They found that dry friction can reverse the direction of motion of the piston which causes a discontinuity or a cusplike singularity in the velocity distribution functions of the piston. Kanazawa {\it et al.}~\cite{kktstshh} studied a nonlinear Langevin equation with non-Gaussian white noise. They obtained an asymptotic formula of the steady distribution function for a large friction coefficient.

Thus, the solid friction between the individual grains and the confinements is of considerable importance in granular flow. One of the most important unsolved problems in the field of dense granular flow is to obtain well-defined hydrodynamic equations. In this context, the formation of plugs in the presence of solid friction causes complications~\cite{rbms10,msrb11}. Also, most of the studies mentioned above are in one dimension. However, in reality, the frictional interaction is present in higher dimensions. Therefore, before tackling the more complicated issues of dense granular flow, we should first address the simplest problem: a single particle affected by noise and solid friction in higher dimensions. In this paper, we analytically and numerically obtain the velocity distribution function, mean-squared velocity in the steady state, velocity autocorrelation function, and diffusion coefficient in $d>1$ for the above problem.

The organization of the paper is as follows. In Sec.~\ref{10sec2}, we describe the details of the model and analytical results. Numerical results are discussed in Sec.~\ref{10sec3}. Finally, we summarize our results in Sec.~\ref{10sec4}.

\section{Details of Model and Analytical Results}\label{10sec2}
The model for single particle Brownian motion in the presence of solid friction and noise was introduced some years ago independently by de Gennes~\cite{pgdg2005} and Hayakawa~\cite{haya2005}. The model consists of a Langevin equation of the form,
\begin{eqnarray}
\label{10eq1}
m\dot{\vec v} = -\mu |\vec N|\frac{\vec v}{|\vec v|} + \vec \xi(t),
\end{eqnarray}
where $m$ is the mass of the particle, $\vec v$ its velocity, $\mu$ the dynamic friction coefficient, between the particle and the $d$-dimensional system against which it rubs, $|\vec N|$ is the normal persistent force applied by that system. In Eq.~(\ref{10eq1}), $\vec \xi(t)$ is a random force applied by the system to the particle due to its vibration (or its thermal agitation, when the particle is extremely small~\cite{pgdg2005}). The average random force is assumed to obey,
\begin{eqnarray}
\label{10eq2}
\langle\vec \xi(t)\rangle=0 ~~~~\text{and}~~~~ \langle\xi_i(t)\xi_j(t') \rangle=\gamma\delta_{ij}\delta(t-t'),
\end{eqnarray}
where $i$ and $j$ denote Cartesian components. (Both authors~\cite{pgdg2005,haya2005} consider slightly more general forces on the left hand side of Eq.~(\ref{10eq1}) but in both cases solid friction and noise play the dominant role.) The most realistic physical realization of Eq.~(\ref{10eq1}) is the two-dimensional system, where the particle is placed on a vibrating plane. The three-dimensional realization is less clear, because the motion of an intruder into a piece of dense material like wood or concrete, for example, will leave an erratic bore in the material and the effect of the particle meeting the bore again must be taken into account. An intruder in the dense sand, for example, might be perhaps a possible realization as the bore is expected to close, in the wake of the intruder.

One of the most important quantities of physical interest is the diffusion coefficient $D$. The expression for $D$ in $d$ dimensions is given by
\begin{eqnarray}
\label{10eqq1}
D = \frac{1}{d}\int_{0}^{\infty}\left\langle\vec v(0)\cdot\vec v(t)\right\rangle dt.
\end{eqnarray}
The quantity $\left\langle\vec v(0)\cdot\vec v(t)\right\rangle$ on the RHS of Eq.~(\ref{10eqq1}) is known as the velocity autocorrelation function (VACF). The detailed calculation of the VACF and $D$ for the problem described by Eq.~(\ref{10eq1}) are given later.

To proceed with the canonical form of the Langevin equation in Eq.~(\ref{10eq1}), we rescale the velocity and time as 
\begin{align}
\label{10eqq2}
\vec v &= \frac{\sigma}{m\mu|\vec N|}\vec v^{\,\prime}, \\
\label{10eqq3}
t &= \frac{\sigma}{\mu^2|\vec N|^2}t^{\,\prime},
\end{align}
where $\vec v^{\,\prime}$ and $t^{\,\prime}$ are, respectively, the dimensionless velocity and time. This results in the canonical form (dropping the primes),
\begin{eqnarray}
\label{10eq3}
\dot{\vec v} = -\frac{\vec v}{|\vec v|} + \vec \chi(t),
\end{eqnarray}
where all the quantities are now dimensionless. The correlations of the rescaled noise are given by $\langle \chi_i(t)\chi_j(t^\prime)\rangle = \delta_{ij}\delta(t-t^\prime)$. The resulting FP equation for the velocity distribution, $P(\vec v)$, is
\begin{eqnarray}
\label{10eq4}
\frac{\partial P}{\partial t} = \vec \nabla_v\cdot\left[\frac{1}{2}\vec \nabla_v + \frac{\vec v}{|\vec v|} \right]P,
\end{eqnarray}
where the gradient is with respect to $\vec v$. It is obvious from the above that the normalized steady state $\left(\frac{\partial P}{\partial t}=0\right)$ is
\begin{eqnarray}
\label{10eq5}
P_s(v) = \frac{2^{d-1}\Gamma(d/2)}{\pi^{d/2}\Gamma(d)}\exp(-2|v|).
\end{eqnarray}
In $d$ dimensions, the mean-squared velocity at steady state is easily obtained from the above,
\begin{align}
\label{10eq6}
\langle v^2 \rangle_s&=\int d\vec{v} v^2 P_s(v)\nonumber\\ &=\frac{1}{4}d(d+1).
\end{align}

To continue to the steady state velocity autocorrelation, we first present the general framework for calculating the time-dependent correlation function. We start by addressing the meaning of a general time-dependent steady state correlation of the form $\langle A(0)B(t)\rangle$ with $t\ge 0$. In the steady state, we measure a function of the dynamical variables $A$. Consequently, the system is not anymore in steady state. After time $t$, during which the distribution is decaying towards a steady state, we measure the quantity $B$ which is another function of the dynamical variables. We take the product and then average. Denoting the dynamical variables by $\{x_i\}$, we can express the required average in the following way,
\begin{eqnarray}
\label{10eq7}
\langle A(0)B(t)\rangle=\int\prod dx_i^{(1)}dx_i^{(2)} P_s\left\{x_l^{(1)} \right\} P\left\{x_l^{(2)},t;x_l^{(1)}\right\} A\left\{x_l^{(1)}\right\}B\left\{x_l^{(2)}\right\},
\end{eqnarray}
where $P_s\left\{x_l^{(1)}\right\}$ is the normalized steady state solution of the Fokker-Planck equation. $P\left\{x_l^{(2)},t; x_l^{(1)}\right\}$ is the distribution function of the dynamical variables $\left\{x_l^{(2)}\right\}$ in which the set $\left\{x_l^{(1)}\right\}$ plays the role of a set of parameters. The distribution $P\left\{x_l^{(2)},t;x_l^{(1)}\right\}$ is the solution of the FP equation for the distribution of the variable $\left\{x_l^{(2)}\right\}$ at time $t$ with the initial condition
\begin{eqnarray}
\label{10eq8}
P\{x_l^{(2)},t;x_l^{(1)}\}=\prod_l \delta\left(x_l^{(2)}-x_l^{(1)}\right).
\end{eqnarray}
When the FP equation can be put in the form,
\begin{eqnarray}
\label{10eq9}
\frac{\partial P}{\partial t}=\sum_i\frac{\partial}{\partial x_i}\left[\frac{1}{2}\frac{\partial}{\partial x_i} + \frac{\partial W}{\partial x_i} \right]P,
\end{eqnarray}
we can obtain a more useful expression for the steady state time-dependent correlation by transforming from the above FP equation to an imaginary-time Schr\"{o}dinger equation, by using the standard transformation,
\begin{eqnarray}
\label{10eq10}
P = P_s^{1/2}\psi=\exp(-W)\psi.
\end{eqnarray}
The imaginary-time Schr\"{o}dinger equation has the form
\begin{eqnarray}
\label{10eq11}
\frac{\partial}{\partial t}\psi = -H\psi,
\end{eqnarray}
where 
\begin{eqnarray}
\label{10eq12}
H = \sum_i\frac{1}{2}\left[\frac{\partial}{\partial x_i} - \frac{\partial W}{\partial x_i}\right]\left[\frac{\partial}{\partial x_i} + \frac{\partial W}{\partial x_i}\right].
\end{eqnarray}
It is clear from the above equation that $H$ is Hermitian, non-negative definite with only a single eigenstate with eigenvalue zero, the ground state exp$(-W)$.

Equations (\ref{10eq10})-(\ref{10eq12}) will enable us to write the required time-dependent correlation in terms of the eigenstates and eigenvalues of $H$. The next step is to express $P\left\{x_l^{(2)}, t;x_l^{(1)}\right\}$ in that way. We note first that
\begin{eqnarray}
\label{10eq13}
\psi\left\{x_l^{(2)}, 0;x_l^{(1)}\right\}=P_s^{-1/2}\left\{x_l^{(1)}\right\}\prod_l\delta\left(x_l^{(2)}-x_l^{(1)}\right),
\end{eqnarray}
where the prefactor of the product of delta functions should have been taken as a function of the $x_l^{(2)}$'s but we are allowed to take it to be the same function of the $x_l^{(1)}$'s, because of the product of the $\delta$-functions multiplying it. The product $\prod_l\delta\left(x_l^{(2)}-x_l^{(1)}\right)$ can be expressed in terms of the eigenstates of $H$, that form a complete orthonormal set as $\sum_n \phi_n^*\left\{x_l^{(1)}\right\}\phi_n\left\{x_l^{(2)}\right\}$, where the $\phi_n$ is a normalized eigenstate of $H$ with eigenvalue $\lambda_n$. Thus,
\begin{align}
\label{10eq14}
P\{x_l^{(2)},t;x_l^{(1)}\}=P_s^{1/2}\left\{x_l^{(2)}\right\}P_s^{-1/2}\left\{x_l^{(1)}\right\}\sum_n \phi_n^*\left\{x_l^{(1)}\right\}\exp(-\lambda_nt)\phi_n\left\{x_l^{(2)}\right\}.
\end{align}
This enables us immediately, by using equations~(\ref{10eq7}) and (\ref{10eq14}), to write the required time-dependent correlation in bra-ket notation,
\begin{eqnarray}
\label{10eq15}
\langle A(0)B(t)\rangle=\sum_n \langle G|A|\phi_n\rangle\exp(-\lambda_nt)\langle\phi_n|B|G\rangle ~~~~ \text{for $t>0$,}
\end{eqnarray}
where $|G\rangle$ is the ground state of the Hamiltonian, which in the coordinate representation is just (the normalized) exp$(-W)$. As we shall see in our specific problem below, the autocorrelation function in dimensions higher than one will be obtained as an approximate finite series. The fact that for an autocorrelation $(B=A)$ all the terms on the right hand side of the above, are positive enables to write down an exact inequality, which facilitates to estimate the error when truncating the expression on the right hand side of Eq.~(\ref{10eq15}).

Let $f(t)\equiv\langle A(0)A(t)\rangle$. We define the $N^{\rm th}$ order approximation of the autocorrelation function,
\begin{eqnarray}
\label{10eq16}
f(t)_N = \sum_{n(\lambda_n\le\lambda_N)}\left|\langle\phi_n|A|G\rangle\right|^2\exp(-\lambda_nt),
\end{eqnarray}
which clearly obeys
\begin{eqnarray}
\label{10eq17}
f(t)_{N-1} + \left[f(0) - f(0)_{N-1}\right]\exp(-\lambda_nt)>f(t)>f(t)_N.
\end{eqnarray}
The above is practical, because in many cases, like in the cases of the velocity correlation we are discussing in this article, $f(0)$ is most easily obtained.

Now we return to our problem. The dynamical variables are the component of the velocity vector and from Eq.~(\ref{10eq4}), it is clear that the corresponding ``classical potential" is $W=|\vec v|$.

We start with the one-dimensional problem, although it was solved more than a decade ago by de Gennes~\cite{pgdg2005}, because our solution is a demonstration of the general method, described by Eq.~(\ref{10eq15}), which we will apply later to higher dimensions. The one-dimensional Hamiltonian is [using Eq.~(\ref{10eq12})],
\begin{eqnarray}
\label{10eq18}
H=-\frac{1}{2}\frac{\partial^2}{\partial v^2} + \frac{1}{2} - \delta(v).
\end{eqnarray}
It is well known that the one-dimensional Hamiltonian with an attractive $\delta$-function potential has a single bound state, exp($-|v|$), which is already normalized. The ($\delta$-function normalized) states in the continuum for which the matrix element of $v$ with the ground state are not zero, are antisymmetric eigenstates of the Hamiltonian, $\frac{1}{\sqrt{\pi}}\sin(kv)$ for positive $k$, with eigenvalues $\lambda_k=\frac{1}{2} + \frac{k^2}{2}$. The matrix elements are easily calculated and the final result is,
\begin{eqnarray}
\label{10eq19}
\langle v(0)v(t)\rangle=\frac{16}{\pi}\exp\left(-\frac{t}{2}\right)\int_0^{\infty}dk \frac{k^2}{[1+k^2]^4}\exp\left[-\frac{k^2}{2}t\right],
\end{eqnarray}
which for large $t$ behaves as $t^{-3/2}\exp\left(-\frac{t}{2}\right)$. For early times, the autocorrelation function reduces to
\begin{eqnarray}
\label{10eq20}
\langle v(0)v(t)\rangle=\frac{1}{16}\exp\left(-\frac{t}{2}\right)(8-4t+5t^2) + O(t^3).
\end{eqnarray}

The one-dimensional result presented above has already obtained by de Gennes\cite{pgdg2005}. However, our above expression corrects certain typographical errors in the de Gennes paper~\cite{pgdg2005}. Concurrently with de Gennes, Hayakawa investigated the mathematical properties of Langevin equation with solid friction~\cite{haya2005}. He obtained the steady state velocity distribution function under a uniform external field. Baule~\textit{et al.}~\cite{baul2010,baul2011} studied the sliding or slipping and stick-slip motion using the path integral formalism in one dimension. Just~\textit{et al.}~\cite{htwj2010} studied a Langevin equation of an object subjected to a viscous drag, an external force, and a Coulomb type friction in one dimension. Menzel and Goldenfeld~\cite{amng2011} studied the FP equation in the presence of both solid friction and viscous friction in $d = 1$. The eigenvalue analysis of their FP equation reduces to a quantum mechanical harmonic oscillator in the presence of a delta potential. Later, Menzel~\cite{menzel15} studied the FP equation in the presence of nonlinear friction and drift using the same technique. In their study, the Hamiltonian corresponding to the FP equation is equivalent to the Hamiltonian of a quantum particle in a box. So, obviously, the more interesting systems to study are the higher-dimensional systems.

\begin{table}
	\centering
	\begin{tabular}{| p{3.8cm} | p{5.8cm} | p{3.5cm} |}
		\hline  
		\hskip 1.0 cm Eigenstate & \hskip 1.2 cm Normalized wave function & \hskip 1.2 cm Eigenvalue \\ [1ex] \hline
		\begin{center}
			Ground State
		\end{center}
		& \begin{eqnarray}
		\phi_0 = \sqrt{\frac{2}{\pi}}~e^{-v}\nonumber
		\end{eqnarray}
		& \begin{align}
		\lambda_0=0\nonumber,\text{not relevant}
		\end{align} \\ \hline
		\begin{center}
			$R_1(v)$
		\end{center}
		& \begin{eqnarray}
		\phi_1 = \frac{1}{\sqrt{6\pi}}\left(\frac{2}{3}\right)^2~ve^{-v/3} \nonumber
		\end{eqnarray}
		& \begin{eqnarray}
		\lambda_1 = \frac{4}{9}\nonumber
		\end{eqnarray} \\ \hline
		\begin{center}
			$R_2(v)$
		\end{center}
		& \begin{align}
		\phi_2 = \sqrt{\frac{3}{\pi}\frac{2^3}{5^5}}~\left(v -\frac{2}{15}v^2\right)e^{-v/5} \nonumber
		\end{align}
		& \begin{eqnarray}
		\lambda_2 = \frac{12}{25}\nonumber
		\end{eqnarray} \\ \hline
	\end{tabular}
	\caption{{\fontfamily{lmss} \fontsize{11}{11}\selectfont Eigenstates and eigenvalues for two dimensions.}}
	\label{10table:1}
\end{table}

In $d$ dimensions, the ``quantum Hamiltonian" is given by,
\begin{eqnarray}
\label{eq21}
H = -\frac{1}{2}\nabla_v^2 + \frac{1}{2} - \frac{(d-1)}{2}\frac{1}{|\vec v|}.
\end{eqnarray}
Since the most important physical dimension for our discussion is $d=2$ and perhaps also $d=3$, we will concentrate here only on these dimensions.

For the two-dimensional case, we note that $\langle\vec v(0)\cdot\vec v(t)\rangle=2\langle v_x(0)v_x(t)\rangle$, where the subscript $x$ denotes the $x$ Cartesian component of the vector. To calculate the latter autocorrelation function, we use the fact that $v_x=|\vec v|\cos\varphi$. This implies at once that only the eigenstates of $H$ which have the form $\phi_n=R_n(v)\cos\varphi$ (we use $v$ instead of $|\vec v|$ to simplify the notation) will have nonvanishing matrix elements $\langle\phi_n|v_x|G\rangle$. The radial equation can be readily solved just as for the $3d$, ``hydrogen atom" problem by the ansatz $R_n(v)=p_n(v)\exp(-\alpha_nv)$, where $n\ge1$ and $p_n$ is a polynomial of degree $n$. We present in the following the long-time behavior of the autocorrelation function by considering the two lowest eigenvalues entering Eq.~(\ref{10eq15}). The eigenstates entering the matrix elements in two dimensions are given in the Table~\ref{10table:1}. The matrix elements are now readily evaluated and the long-time autocorrelation function is
\begin{eqnarray}
\label{10eq22}
\langle\vec v(0)\cdot\vec v(t)\rangle = \frac{3^5}{2^9}\exp\left(-\frac{4t}{9}\right) + \frac{5^5}{2\times3^9}\exp\left(-\frac{12t}{25}\right).
\end{eqnarray}

For three-dimensional case, we take $\langle\vec v(0)\cdot\vec v(t)\rangle=3\langle v_z(0)v_z(t)\rangle$, where $v_z$ denotes the $z$ component of $\vec v$, which equals $v\cos\theta$. The only states to be used for the calculation for the matrix elements of $v_z$ are the ground state and states of the form $R_n(v)\cos\theta$. The normalized ground state and the two states of the required form with the lowest eigenvalues are summarized in the Table~\ref{10table:2}.
\begin{table}
	\centering
	\begin{tabular}{| p{3.8cm} | p{5.8cm} | p{3.5cm} |}
		\hline  
		\hskip 1.0 cm Eigenstate & \hskip 1.2 cm Normalized wave function & \hskip 1.2 cm Eigenvalue \\ [1ex] \hline
		\begin{center}
			Ground State
		\end{center}
		& \begin{eqnarray}
		\phi_0 = \sqrt{\frac{2}{\pi}}~e^{-v}\nonumber
		\end{eqnarray}
		& \begin{align}
		\lambda_0=0\nonumber,\text{not relevant}
		\end{align} \\ \hline
		\begin{center}
			$R_1(v)$
		\end{center}
		& \begin{align}
		\phi_1 = \frac{1}{4\sqrt{2\pi}}~ve^{-v/2} \nonumber
		\end{align}
		& \begin{eqnarray}
		\lambda_1 = \frac{3}{8}\nonumber
		\end{eqnarray} \\ \hline
		\begin{center}
			$R_2(v)$
		\end{center}
		& \begin{align}
		\phi_2 = \sqrt{\frac{1}{2\pi}}~\frac{2^2}{3^3}~\left(1 - \frac{v}{6}\right)ve^{-v/3} \nonumber
		\end{align}
		& \begin{align}
		\lambda_2 = \frac{4}{9}\nonumber
		\end{align} \\ \hline
	\end{tabular}
	\caption{{\fontfamily{lmss} \fontsize{11}{11}\selectfont Eigenstates and eigenvalues for three dimensions.}}
	\label{10table:2}
\end{table}
The corresponding long-time $3d$ autocorrelation function is given by,
\begin{eqnarray}
\label{10eq23}
\langle\vec v(0)\cdot\vec v(t)\rangle = \frac{2^{15}}{3^9}\exp\left(-\frac{3t}{8}\right) + \frac{3^7}{2^{13}}\exp\left(-\frac{4t}{9}\right).
\end{eqnarray}

A number of remarks are in order here. Since the method of obtaining the relevant bound states in any number of dimensions and the corresponding eigenvalues is straightforward, it is possible to improve the expression to any required accuracy. It has to be recalled though, that we need a complete set of eigenstates and therefore, like in one dimension, there is also a contribution of the states in the continuum (states with eigenvalues of $H$ larger than $1/2$), which has to be taken into account.

Now we return to the calculation of the diffusion coefficient $D$ in Eq.~(\ref{10eqq1}), which requires the VACF. Since all the VACFs are calculated in dimensionless form, we will rewrite Eq.~(\ref{10eqq1}) in dimensionless units. Using Eqs.~(\ref{10eqq2}) and (\ref{10eqq3}), $D$ in Eq.~(\ref{10eqq1}) reduces to (dropping primes of $\vec v^{\,\prime}$ and $t^\prime$)
\begin{eqnarray}
\label{10eq24}
D = \frac{1}{d}\left(\frac{\gamma}{m\mu|\vec N|}\right)^2\left(\frac{\gamma}{\mu^2|\vec N|^2}\right)\int_{0}^{\infty}\left\langle\vec v(0)\cdot\vec v(t)\right\rangle dt.
\end{eqnarray}
Clearly, the integral on the right-hand side of Eq.~(\ref{10eq24}) is a constant, say $K_d$. This can be evaluated numerically from the corresponding integral expressions for the VACF. Therefore, Eq.~(\ref{10eq24}) yields
\begin{eqnarray}
\label{10eq25}
D = \frac{K_d}{d}\frac{\gamma^3}{m^2\mu^4|\vec N|^4}.
\end{eqnarray}
Moreover, one can define the granular temperature, $T$ from the mean kinetic energy as
\begin{eqnarray}
\label{10eq26}
T \sim \int_{-\infty}^{\infty} \vec v^{\,2}P(\vec v)d\vec v.
\end{eqnarray}
Using the dimensionless form of $\vec v$ from Eq.~(\ref{10eqq2}) in Eq.~(\ref{10eq26}), we obtain $T\propto\gamma^2$. Combining this with Eq.~(\ref{10eq25}), we find $D\propto T^{3/2}$, in contrast to $D\propto T$ for usual Brownian motion~\cite{balki}.

\section{Numerical Results}\label{10sec3}
In this section, we solve numerically Eq.~(\ref{10eq3}) to obtain the $\left<v^2\right>(t)$ and $\left<\vec v(0)\cdot\vec v(t)\right>$ in $d=1, 2~\text{and}~3$ respectively, using Euler-discretization scheme~\cite{pkep1992}. The discretized version of Eq.~(\ref{10eq3}) is as follows
\begin{eqnarray}
\label{10eq24x}
v(t+\Delta t) = v(t) - {\rm sgn}[v(t)]\Delta t + \sqrt{\Delta t}~n(t)
\end{eqnarray}
with $\left<n(t)\right>=0$ and $\left<n(t)n(t')\right>=\delta_{tt'}$ in $d=1$. Further,
\begin{eqnarray}
\label{10eq25x}
v_i(t+\Delta t) = v_i(t) - \frac{v_i(t)}{|\vec v(t)|}\Delta t + \sqrt{\Delta t}~n_i(t)
\end{eqnarray}
\begin{figure}
	\centering
	\includegraphics[width=0.80\textwidth]{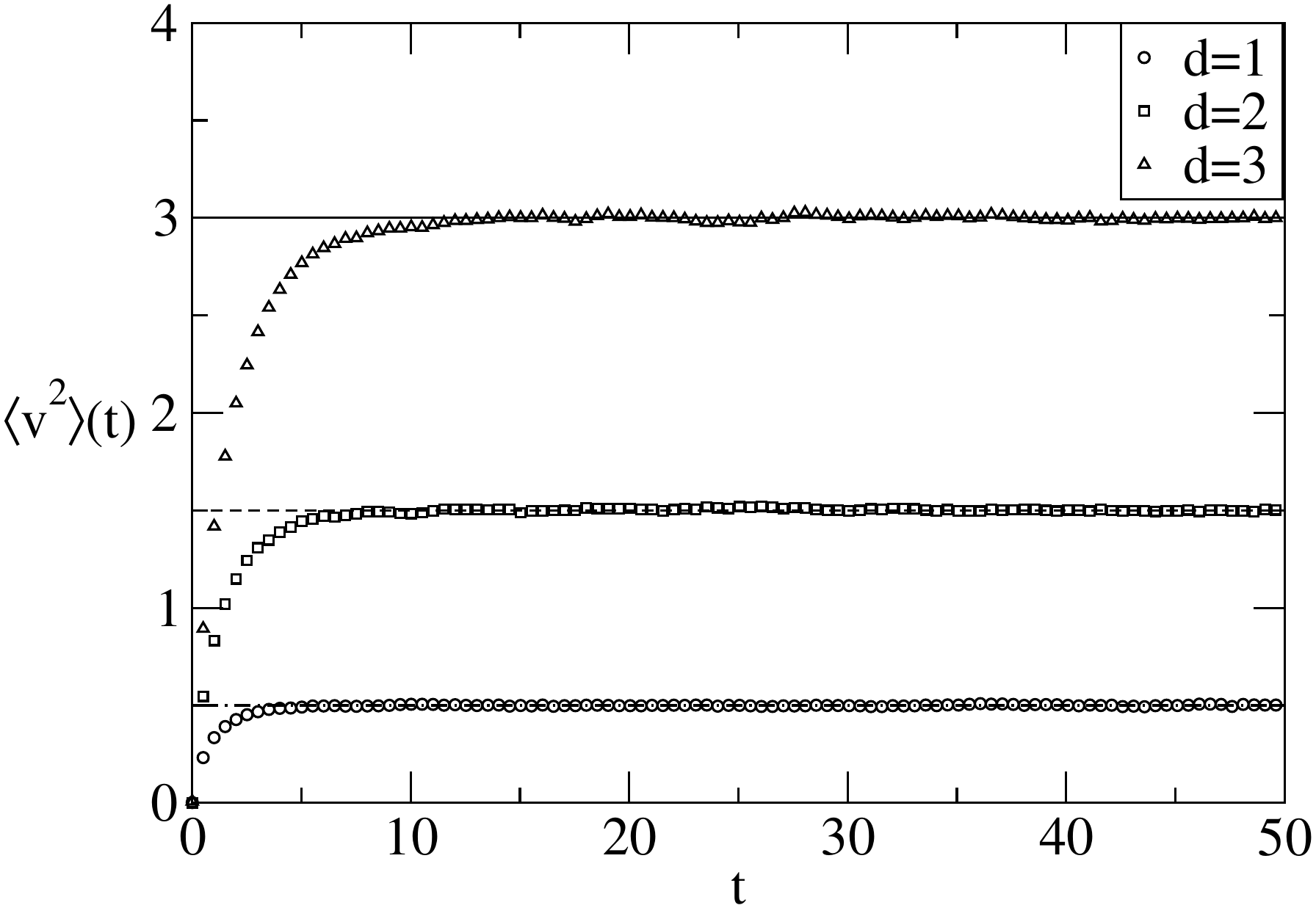}
	\caption{\label{10fn1} {\fontfamily{lmss} \fontsize{11}{11}\selectfont Time evolution of \textit{$\left<v^2\right>(t)$} for different spatial dimensions, $d$, as mentioned. Numerical data is obtained by averaging over $N_p=10^5$ single particles. Solid lines represented the analytically obtained value of $\left<v^2\right>_s$ for different value of $d$. Clearly, our analytical results are in good agreement with numerical results.}}
\end{figure}
with $\left<n_i(t)\right>=0$ and $\left<n_i(t)n_j(t')\right>=\delta_{ij}\delta_{tt'}$ in $d>1$ and $i=x,y,z$. Here $\Delta t$ is the discretized time step, and $n(t)$ is the Gaussian white noise of unit width. In the simulation, the numerical value of $\Delta t=0.001$ is used and the results are obtained by averaging over $10^5$ trajectories. At $t=0$, we start with zero initial velocity of the particle. The velocity of each particles is updated according to Eq.~(\ref{10eq24x}) in $d=1$ and Eq.~(\ref{10eq25x}) in $d>1$. In Fig.~\ref{10fn1}, we plot $\left<v^2\right>(t)$ in different spatial dimensions. Numerical details are given the figure caption. Clearly, in the steady state, the numerical value of $\left\langle v^2\right\rangle(t)$ agrees well with the analytically obtained $\left<v^2\right>_s$ in Eq.~(\ref{10eq6}) for all physical dimensions.

\begin{figure}
	\centering
	\includegraphics[width=0.95\textwidth]{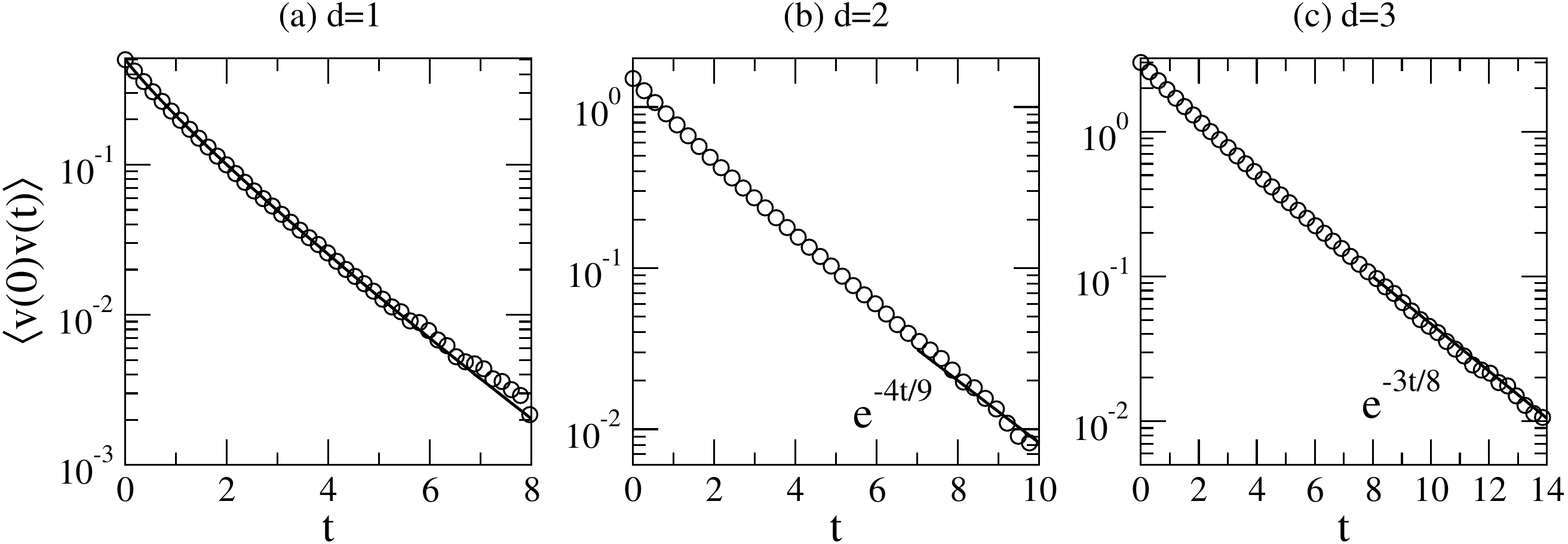}
	\caption{\label{10fn2} {\fontfamily{lmss} \fontsize{11}{11}\selectfont Plot of the velocity autocorrelation function (VACF) vs. $t$ in the steady state on linear-log scale. The numerical results are shown by open circles. (a) The VACF in $d=1$. Solid line represents the result obtained by exact VACF, given by Eq.~(\ref{10eq19}). (b) The VACF in $d=2$. At early times, VACF decays as $e^{-t/2}$. At later times, the VACF decays as $e^{-4t/9}$, shown by the solid line. (c) The VACF in $d=3$. In the early stage, the decay of VACF is similar to the case with $d=2$. However, at the late stage, VACF decays as $e^{-3t/8}$ which is shown by the solid line.}}
\end{figure}

Figure~\ref{10fn2} shows the velocity autocorrelation function (VACF) in the steady state. Starting from the initial condition, the velocity of individual particle is updated up to $t=20$, when $\left<v^2\right>(t)$ reaches steady state. Figure~\ref{10fn2}(a) shows the VACF in one dimension. The numerical results obtained from the simulation follows the exact analytical VACF given by a numerical integration of Eq.~(\ref{10eq19}). For $d>1$, we have approximate solutions for VACF. In Fig.~\ref{10fn2}(b), the VACF in $d=2$ shows a crossover from exp$(-t/2)$ at early times to exp$(-4t/9)$ at later times in agreement with the analytical predictions, given by Eq.~(\ref{10eq22}). Of course, it is not easy to numerically distinguish between the early-time exp$(-t/2)$ and the late-time exp$(-4t/9)$ behavior. The early-time behavior of the VACF in $d=3$ is similar to that of in $d=2$ as shown in Fig.~\ref{10fn2}(c). However, in the late stage, VACF decays as exp$(-3t/8)$ as given by Eq.~(\ref{10eq23}).

\section{Summary}\label{10sec4}
Let us conclude our results presented here with a summary and discussion. Here, we studied the Brownian motion of a solid particle on a vibrating plate. Solid friction models the interaction between two solid surfaces, which is proportional to the direction of the relative velocity between two interacting surfaces. Gaussian white noise models the vibration of the plate. We obtain the steady state velocity distribution function and mean-squared velocity in $d$ dimensions. The numerical results for mean-squared velocity are in good agreement with the analytical results. We also analytically calculate the velocity autocorrelation function (VACF) up to $d=3$ dimensions. In $d=1$, we have the exact form of the VACF followed by the numerical result. However, for $d>1$, we obtain an approximate expression for the VACF. The numerical results show that the VACF in the early stage decays as exp$(-t/2)$ for both $d=2$ and $d=3$; and shows crossover to exp$(-4t/9)$ for $d=2$ and exp$(-3t/8)$ for $d=3$ in the late stage which are in agreement with the analytical results. Next, we calculate the diffusion coefficient, $D$ which depends on $\mu$ and $\gamma$. We believe that the results presented here will be useful for the further study of rheology of dense granular matter. 

\section{Acknowledgments}
PD acknowledges financial support from CSIR, India. The research of MS, grant number 839/14, was supported by the ISF within the ISF-UGC joint research program framework. SP is grateful to UGC, India for support through an Indo-Israeli joint project. He is also grateful to DST, India for support through a J. C. Bose fellowship.

\end{document}